\documentclass[aps,prb,showpacs,twocolumn]{revtex4}
\usepackage{amsmath}
\usepackage{amssymb}
\usepackage{epsfig}
\begin{document}
\title{Noise in superconductor-quantum dot-normal metal structures in the Kondo regime}
\author{P. Devillard$^{1,2}$ and A. Cr\'epieux$^{1,3}$}
\affiliation{$^1$ Centre de Physique Th\'eorique de Marseille
        CPT, case 907, 13288 Marseille Cedex 9 France}
\affiliation{$^2$ Universit\'e de Provence, 3, Place Victor Hugo, 
13331 Marseille cedex 03, France}
\affiliation{$^3$ Universit\'e de la M\'editerran\'ee, 13288 Marseille Cedex 9, France}
\begin{abstract}
We consider a N-dot-S junction in the
 Kondo regime in the limit where the superconducting gap is much smaller
 than the Kondo temperature. A generalization of the floating of the Kondo resonance
 is proposed and many body corrections to the average subgap current are calculated.
 The zero frequency noise is computed 
and the Fano factor sticks to the value $10/3$ for all voltages
 below the gap. Implications
 for finite frequency noise are briefly discussed.
\end{abstract}
\pacs{71.10.Ay, 71.27.+a, 72.15.Qm, 73.63.Fg} 
\maketitle

Possible realizations of quantum dots have revived the interest 
 in Kondo physics \cite{GlazmanPustilnik,Hewson}. 
 For normal electrodes, the dot spin is screened and the system behaves in first approximation
 as a resonant level at the Fermi energy. However, this picture is not sufficient; the
system is a Fermi liquid with interactions, morevover, the ratio 
of elastic and inelastic backscattering obeys some universality\cite{Nozieres}, 
which can be obtained from the concept of floating of the Kondo resonance. 
Current noise is a useful tool to obtain informations on interactions, 
which are not present in the average current $\langle I \rangle$. Recently,
 the zero frequency noise $S$ and the Fano factor $F=S/2e \langle I \rangle$
 in the SU(2) case\cite{SelavonOppen} and the SU(4) 
case\cite{MoraLeyronasRegnault,VitushinskyClerkleHur} were calculated 
and confirmed experimentally\cite{Heiblumgroup} for SU(2).

What happens to this effect in the superconducting case? 
This poses the problem of interplay between Kondo physics and superconductivity,
 already present in heavy fermion compounds\cite{Hewson} and underdoped high-$T_c$ materials
\cite{Anderson}. Devices with a dot between two
 superconducting electrodes have been extensively studied both
 theoretically 
\cite{YoshiokaOhashi,AvishaiGZ,
KangChoiBelzig,SianoEgger,LevyYeyati,Meden,MLeeJonckMart} and experimentally 
\cite{Wernsdorfergroup,DeblockBouchiat} with emphasis on the Josephson current, 
and on how increasing
 the gap $\Delta$ destroys the Kondo effect.
 Here, we study the noise of the subgap Andreev 
current in a normal metal-dot-superconductor structure for $\Delta \ll T_K$. 
While in this regime, there is still no destruction of the Kondo resonance 
by the presence of the 
gap $\Delta$\cite{GlazmanPustilnik,KangChoiBelzig}, the interplay between one and many-particle scattering, 
Andreev and normal reflexion, is far from trivial. 
The central result of this paper is a generalization of the floating of the Kondo resonance  
 in the case where one electrode is superconducting. The most noticeable consequence
 is a constant Fano factor, equal to
 $10/3$ for all voltages bias below the gap. This could be tested experimentally on carbon 
nanotubes.          

{\it Model:} A quantum dot with effectively one level
 of energy $\epsilon_0$ is placed between two electrodes, 
the left one being normal and the right one being
 a usual BCS superconductor, see Fig. 1.
 The on site repulsion $U$ on the dot is supposed to be the largest energy of the problem. 
Electrons can hop from the lead to the dot with amplitude $\tau$, implying a broadening
 of the level $\Gamma = 2 \pi \rho(\epsilon_F) \vert \tau \vert^2$, with
 $\rho(\epsilon_F)$ the density of states at the Fermi energy in the normal metal. 
We abide by the particle-hole symmetric case, 
for which $\epsilon_0 = - \,U /2$. The
 Hamiltonian reads
\begin{eqnarray}
H  \, = \, \sum_{k,\sigma,p} 
(\epsilon_{k} - \mu_p)
c^{\dagger}_{k,\sigma,p} c_{k,\sigma,p}
\, + \tau c_{k,\sigma,p} d_{\sigma}^{\dagger} + h.c. \nonumber \\
+ \sum_{k} \Delta c^{\dagger}_{k,\uparrow,R}
 c^{\dagger}_{-k,\downarrow,R} + h.c.
+ \epsilon_0 (n_{\uparrow} + n_{\downarrow}) + \, U n_{\uparrow}n_{\downarrow},
\end{eqnarray}
where $\Delta$ is the superconducting order 
parameter; $p=L$ for the normal lead on the left 
and $p=R$ for the superconducting one; the operators
 $c^{\dagger}_{k,p,\sigma}$ and $d_{\sigma}^{\dagger}$
create an electron on the $p$ lead and on the dot,
 respectively and $n_{\sigma}= d_{\sigma}^{\dagger} d_{\sigma}$. The chemical potentials are 
$\mu_L=eV$ and $\mu_R=0$, with $V$ the voltage bias. For
 $\Delta \, \ll \, T_K$, and even when both electrodes are
 superconducting,  it was shown by various methods \cite{GlazmanPustilnik},
 both analytical and numerical \cite{KangChoiBelzig,SianoEgger}
 that the Kondo resonance subsists despite the gap.  
In the non-Kondo case, the average current has been calculated
 in Ref. \onlinecite{AvishaiGZ}. Also, in the low temperature regime, in the 
Kondo case, the average current and the noise were 
estimated by using slave-boson methods but
 this method is an effective one-body approximation to the full Kondo Hamiltonian
 and thus, is not sufficient to capture the complexity of the Kondo
 Hamiltonian, already when both electrodes are normal \cite{SelavonOppen}. 

{\it Method and one-body setting:} Here, we want to generalize the calculation of the Kondo noise 
when two electrodes are normal to the case where one electrode is superconducting. 
\begin{figure}[h] 
\epsfxsize 7. cm  
\centerline{\epsffile{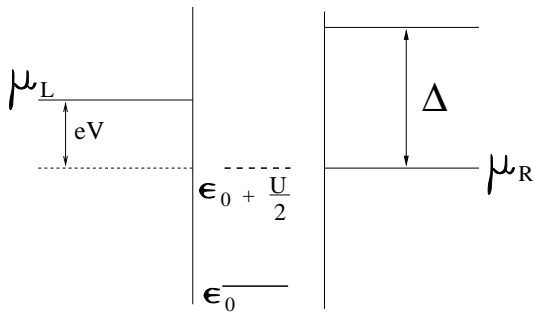}}
\caption{
Quantum dot sandwiched between a normal electrode on the left with chemical potential
 $\mu_L$ an a superconducting one on the right 
with gap $\Delta$ and chemical potential
 $\mu_R=0$. Only one level of negative energy $\epsilon_0$ intervenes. Hopping amplitudes 
on and from the dot to both electrodes are assumed to be equal. The effective energy level 
taking into account the Hartree correction is ${\tilde \epsilon}_0 = 
\epsilon_0 +\, U/2$, in heavy dashed line.}
\end{figure} 
In order to include the many-body effects, we consider a slightly different model for the Kondo dot,
 which has been used for modelling an imperfect NS junction \cite{Beenakker},
 see Fig. 2. The dot is a scatterer
 placed in front of the superconductor and produces dephasing, whereas in reality, the dot acquires
 off-diagonal order
 due to hopping on and from the superconductor. An incident electron 
 of energy $\epsilon$ is scattered by the dot and then 
 Andreev reflected as a hole
 by the NS interface, supposed to be perfect, with a modulus one and 
a phase $-i \, {\rm atan} (\epsilon/\Delta)$,
 and then, this hole gets scattered by the dot, (see Fig. 2).
 The dephasing of the electron by the scatterer 
is $\delta_e = \delta_e^{(0)} + \delta_e^{(1)} + ...$, where
 $\delta_e^{(0)} = - \pi/2$, $\delta_e^{(1)}$ is first order in $\epsilon/T_K$ and the dots
 represent higher orders in $\epsilon/T_K$.
 The energy transmission of electrons through the scatterer is 
$T(\epsilon)= sin^2(\delta_e)$. The same
 thing happens for the holes with dephasing $\delta_h$. 
Using the Bogolubov-de Gennes (BdG) equations
 enables to express the $s-$matrix in Nambu space for the whole structure, 
  which is $4 \times 4$ but, below the gap, reduces to a $2 \times  2$ 
reflexion matrix, because no transmission of one-particle excitation occurs 
in the superconductor. The
 energy dependent normal reflection coefficient $r_N$ and the 
Andreev amplitude reflection coefficient $r_A$
 can be extracted in terms of $\delta_e$  
\begin{figure}[h] 
\epsfxsize 7. cm  
\centerline{\epsffile{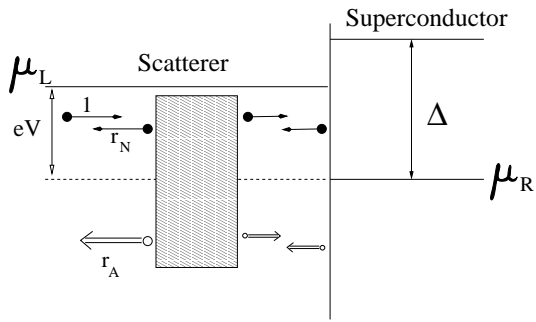}}
\caption{
Simplified model for the $N-dot-S$ system. A scatterer, having no off-diagonal Green's function, is 
 placed in front of the superconductor accounts for the imperfect transmission 
properties of the dot. The interface with the superconducting electrode is taken to be perfect
 so that only Andreev reflexion occurs at the interface and only normal reflexion or transmission occurs
across the scatterer. Proximity effects are neglected. An incident electron is partially 
 transmitted through the scatterer and then totally Andreev reflected as a hole. This hole is in turn
 partially transmitted as a hole but also partially reflected by the scatterer, etc...  
All processes need to be resummed 
to infinite order, the amplitudes $r_N$ and $r_A$ are the result of this (coherent) resummation.
 However, this is equivalent to solving the BdG equations.}
\end{figure} 
and $\delta_h$. 
We want to make contact with the original model of Eq (1).
 If $T_K$ were infinite, then,
 the one-body picture of a resonant model centered 
 at the Fermi level, with width $T_K$ would be valid. We are
 close to the unitary limit and we need the dot Green's function 
in this limit. In the case where both electrodes are normal, the retarded 
Green's function is of the form $z^{-1} \bigl(\epsilon - 
{\tilde \epsilon_0} - 2 i {\tilde \Gamma} sgn(\epsilon) \bigr)^{-1}$ with
 $z=1- (\partial  \Sigma^{ret}(\epsilon) / \partial \epsilon)\vert_{\epsilon=\epsilon_F}$ and
 ${\tilde \Gamma} = z^{-1} \Gamma \simeq T_K/2$; $\Sigma^{ret}(\epsilon)$ 
being the retarded self-energy of the dot and ${\tilde \epsilon}_0=\epsilon_0+U/2$. 
 This has been used 
justified theoretically by Ref. \onlinecite{Hewson}.
When both electrodes are superconducting, for $\Delta \ll T_K$, a similar procedure exists and 
has been used in Ref. \onlinecite{LevyYeyati} to study the dynamics of Andreev states. 
We thus use these Green's functions, adapted to our case, 
and following the same steps as in Ref. \onlinecite{AvishaiGZ},
 the conductance $g_A$ reads
\begin{eqnarray} 
g_A = 4 {e^2 \over h} (eV)^{-1} 
\int_0^{eV} 
\Bigl({\epsilon \over \tilde\Gamma/2}\Bigr)^2
\Bigl({\epsilon \over \Delta}\Bigr)^2  
\Bigr\rbrack \, d \epsilon.
\end{eqnarray} 
On the other hand, the model of Ref. \onlinecite{Beenakker} 
leads to a conductance 
\begin{eqnarray}
g_B =  4 {e^2 \over h} (eV)^{-1} \int_0^{eV} \Bigl({T(\epsilon) \over 2 - T(\epsilon)} \Bigr)^2 \,
 d \epsilon .
\end{eqnarray}
Thus, for $g_A$ and $g_B$ to have the same expression, 
we adjust the dephasing of the scattering 
center to be
 $\delta_e^{(1)} = (\epsilon /\Delta)\,
(\epsilon / {\tilde \Gamma}) \equiv (\epsilon / T_K) \, \alpha_1(\epsilon)$, where
 $\alpha_1$ is a function of $\epsilon$. The same occurs for $\delta_h^{(1)}$. This form does not correspond to a resonant
level at $\epsilon=0$ and width ${\tilde \Gamma}/2$. In the original model,
 the dot acquires some off-diagonal matrix element in Nambu space. Very qualitatively, this favors
 direct Andreev reflexion of a wave incoming on the dot. Since,
for the whole structure,  $\vert r_N \vert^2 + 
 \vert r_A \vert^2 =1$, there will be less normal scattering. The amount of normal scattering
 increases as $(\epsilon- \epsilon_F)^2$ instead of 
$(\epsilon - \epsilon_F)$ for the normal case. Let us denote by 
$G^{ret}_{dot \, 1,1}$ the upper Nambu component of the retarded
 dot Green's function, calculated  without the many-body corrections, 
using the renormalized parameter $\tilde \Gamma$ 
and putting aside the usual multiplicative renormalization  factor $z$. 
For energies higher than the gap, the form of the normal part of the spectral density
 of the dot $\rho_{1,1}(\epsilon)  = - \pi^{-1}\, Im G^{ret}_{dot \, 1,1}$
 is very close
 to the usual shape for the normal case. Surprisingly, this
 persists down to energies barely larger than
 $\Delta$. Of course, this is not the exact $\rho_{1,1}(\epsilon)$ but
 this feature seems to
 be shared by more elaborate numerical solutions for the S-dot-S case, 
such as numerical renormalization group (NRG) \cite{KangChoiBelzig} 
or functional renormalization group (fRG) \cite{Meden,DeblockBouchiat}. 

{\it Many-body calculation:} 
Now, we want to put the many-body corrections since $T_K$ is not infinite. 
The electrons build the Kondo resonance from hopping to the 
 normal lead and also from hopping to the superconductor. Unlike the normal case,
 the argument of doping the normal side to establish the floating of the Kondo 
resonance\cite{Nozieres,MoraLeyronasRegnault}
 does not work directly, because doping the metal and changing
 the chemical potential for the electrons will also
 change it for the holes. However, for $\Delta \ll \epsilon \ll T_K$, 
$\delta_e^{(1)}$ is linear 
in $\epsilon$,
 ($\alpha_1$ becomes energy independent) and the argument can be used.      
Thus we have to cancel the linear contribution in $1/T_K$ of $\delta_e$ by a
 many-body interaction between electrons (and holes), as in usual Fermi-liquid theory.
 The fixed point Hamiltonian is
\begin{eqnarray}
H_{FP} \, = \, H_0 + H_{int},
\end{eqnarray}
with $H_0$ being the one-body
 Hamiltonian, involving scattering states (first order in $1/T_K$) and
$H_{int}$ is the interaction between quasiparticles. The simplest form 
maintaining particle-hole symmetry is
\begin{eqnarray}
H_{{\rm int}} = {1 \over T_K} 
\biggl\lbrack \sum_{k_1,k_2,k_3,k_4} \!\!
 \beta_1(\epsilon_{k_1} +\epsilon_{k_2} + 
\epsilon_{k_3} + \epsilon_{k_4}) \nonumber \\
\times \, \Bigl( b_{k_1,\uparrow}^{\dagger}   b_{k_2,\downarrow}^{\dagger}
  b_{k_3,\downarrow}   b_{k_4,\uparrow}  +
 B_{k_1,\uparrow}^{\dagger}   B_{k_2,\downarrow}^{\dagger} 
 B_{k_3,\downarrow}   B_{k_4,\uparrow}\Bigr) \biggr\rbrack,
\end{eqnarray}
 with $b_{k,\sigma} = (c_{k,\sigma,L} + c_{k,\sigma,R})/\sqrt{2}$ and
$B_{k,\sigma} = (c_{-k,{\overline \sigma},L}^{\dagger} +
 c_{-k,{\overline \sigma},R}^{\dagger})/\sqrt{2}$ with ${\overline \sigma} = - \sigma$ and
$\beta_1(\epsilon)$ is a function of energy. 
Here, a  right-mover state with energy $\epsilon$ will have some normal reflected part. No elastic
 scattering Hamiltonian is necessary \cite{MoraLeyronasRegnault}.

Because of Hartree-Fock corrections to order one in $\beta_1$,
 the dephasing $\delta_e^{(1)}$ will be changed to
 $\delta_e^{(1)} - \, (\beta_1 /T_K)
 \delta n_{\overline \sigma}$, so that $\beta_1 = \alpha_1$ 
to cancel the $1/T_K$ overall contribution.

The main assumption of this paper is that this can be continued all the way 
 down to $\epsilon=0$. Then, $\beta_1(\epsilon)$ will have to
 cancel $\alpha_1(\epsilon)$.       

Now, we calculate the average current in the Keldysh formalism, 
perturbatively to order $1/T_K^2$, following the method of Ref. 
\onlinecite{VitushinskyClerkleHur}.
 The right mover scattering states are thus of the form, in Nambu space,
 far on the left of the scattering center
\begin{eqnarray}
\psi_{{\cal R}}(\epsilon_k,x) = {1 \over \sqrt{k}} \!
\Biggl\lbrack \bigl( e^{ikx} \! + \! r_N e^{-ikx}  \bigr)
\begin{pmatrix} 1 \cr 0
\end{pmatrix}
\! + \! r_A e^{-ikx} \begin{pmatrix} 0 \cr 1
\end{pmatrix} \Biggr\rbrack \!\! .
\end{eqnarray} 
The left-mover scattering state consists in sending a hole from the left, which is partially
 normally reflected and
 also Andreev reflected as a left propagating electron.
 At the one-body level, the
 current operator has two components 
and is then expressed in terms of the BdG wave function,
 and falls into three parts, $I_{R}$, involving
 only right-moving scattering states, 
$I_L$ with solely left-movers and $I_{OD}$ 
which involves both right and left-movers. Each of these
 three terms gives rise to two contributions. One contribution,
 denoted by subscript $1$ comes from the electron part of the two component
 Nambu wave function and the other one from the hole part (subscript $2$).
 For example $I_{R,1} \, = \, \sum_{k, \sigma} (1 - \vert r_N \vert^2) 
\psi_{k,\sigma,R,1}^{\dagger} \psi_{k,\sigma,R,1}$, with $\psi_{k,\sigma,R}$ is the operator creating a two-component Nambu 
 right mover, and $\psi_{k,\sigma,R,1}$ denotes its upper component.
 $r_N$ and $(1 - \vert r_A \vert)$ are
 order $1$ in $1/T_K$. We could work with these scattering states 
 but we prefer to use the zero-th
 order scattering states, (i.e. $r_N=0$ and $\vert r_A \vert = 1$). This is at the price of having to 
introduce an elastic scattering Hamiltonian. The fixed point Hamiltonian is 
$H^{\prime}_{FP} = H^{\prime}_0 + H^{\prime}_{\alpha} + H^{\prime}_{int}$
where $H^{\prime}_0$ is the Hamiltonian for a perfect scatterer 
(zero$^{{\rm th}}$ order in 
 $1/T_K$) and  \
\begin{eqnarray}
H^{\prime}_{\alpha}  = 
 {1 \over \pi \nu T_K} 
\sum_{i=1}^2 
\sum_{k,k^{\prime},\sigma} 
\Bigl\lbrack{\epsilon_{k} + \epsilon_{k^{\prime}} \over
 2}\Bigr\rbrack \, \nonumber \\
\times \, \alpha_2(\epsilon_k + \epsilon_{k^{\prime}})\,
 {\cal B}_{k,\sigma,i}^{\dagger} 
 {\cal B}_{k^{\prime},\sigma,i},
\end{eqnarray}
with ${\cal B}^{\dagger}_{k,\sigma,i} \, = \, (\Phi^{\dagger}_{k,\sigma,R,i}
 + \Phi^{\dagger}_{k,\sigma,L,i})/\sqrt{2},$ where 
$\Phi^{\dagger}_{k,\sigma,R,i}$ are the Nambu components of operators creating a 
scattering state to zero$^{{\rm th}}$ order in $1/T_K$. $\alpha_2$ is 
adjusted so that $H^{\prime}_{\alpha}$ gives the same dephasing
 $\delta_e$ to order $1/T_K$ as $H_0$ (Eq. (4)). $\alpha_2$ is proportionnal to
 $\alpha_1$; we obtain
 $\alpha_2(\epsilon) = A \, \epsilon/\Delta$, with $A=2$. 
The two-body Hamiltonian is
written in the form 
\begin{eqnarray}
H^{\prime}_{{\rm int}} &=& {1 \over \pi \nu T_K}
\sum_{i=1}^2 
\sum_{k_1,k_2,k_3,k_4}
 \beta_2(\epsilon_{k_1}+\epsilon_{k_2}+\epsilon_{k_3}+ \epsilon_{k_4}) \nonumber \\
&\,&  \times \, {\cal B}_{k_1,\uparrow,i}^{\dagger}   {\cal B}_{k_2,\downarrow,i}^{\dagger}
   {\cal B}_{k_3,\downarrow,i}   {\cal B}_{k_4,\uparrow,i},
\end{eqnarray}
$\beta_2$ depends on energy $\epsilon$. 
For the same reason that $\beta_1$ and $\alpha_1$ were not independent,
 $\beta_2$ and $\alpha_2$ are also linked.
 We find $\beta_2(\epsilon) = B \epsilon/\Delta$  with $B=2$. 
For calculating the inelastic part, it suffices to expand to second order in $\beta_2$ 
the quantity $(1/2) \, \sum_{\eta = \pm } 
\langle  T_{{\cal C}} \,I(t^{\eta}) \, 
\exp \Bigl( - i \int_{{\cal C}} H^{\prime}_{{\rm int}}(t^{\eta_1}) dt^{\eta_1} \Bigr)
\rangle$, where $\eta$ and $\eta_1$ are Keldysh indices, ${\cal C}$ denotes
 the Keldysh contour and $T_{{\cal C}}$ the corresponding time ordering.
 Lumping these with the elastic backscattering contribution results in
 the following expression for the averaged current
\begin{figure}[h] 
\epsfxsize 7. cm  
\centerline{\epsffile{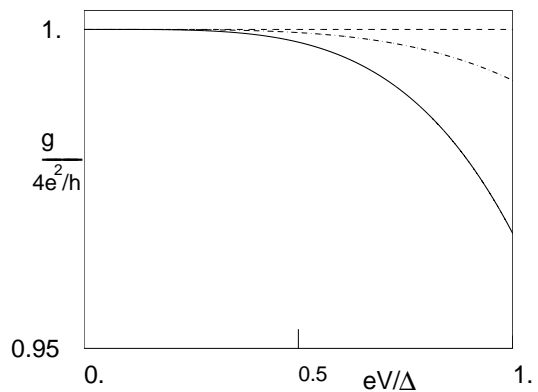}}
\caption{Conductance in units of $4 e^2/h$ vs. $eV/\Delta$ 
for $\Delta/T_K = 0.2$, taking into account
 the many-body terms (solid line), in the resonant
 level approximation (dash-dotted line) and 
for infinite $T_K$ (BTK result), (dashed line).}
\end{figure} 
\begin{eqnarray}
\langle I \rangle \, = \, 4 {e^2 \over h} V \,
\Bigl\lbrack  
1 - 
\Bigl({A^2+ 3 B^2 \over 20}\Bigr)
\Bigl({eV \over \Delta}\Bigr)^2 \Bigl({eV \over T_K}\Bigr)^2
\Bigr\rbrack.
\end{eqnarray}

 As usual, this can be retrieved in a simple way by using Fermi golden rule
 for the elastic process and the three inelastic processes, 
 $({\cal R},{\cal R}) \rightarrow ({\cal R},{\cal L})$, 
 $({\cal R},{\cal R}) \rightarrow ({\cal L},{\cal L})$ and 
 $({\cal R},{\cal L}) \rightarrow ({\cal L},{\cal L})$,
 where ${\cal R}$ is a right-mover and ${\cal L}$ a left-mover. 
For example, for the second process, two right-movers are backscattered;
 one right-mover carries a charge $e$ for the incoming electron and 
 also $e$ from the Andreev reflected hole, so $2e$ in total. The current is of the form
 $(1/2)\, d (N_{{\cal L}} - N_{{\cal R}})/dt$, where
 $N_{{\cal L}\,({\cal R})}$ is
 the operator number for left (right) movers and the change in the current
 is $4e$. The diagram multiplicities are the same as for the
 normal case but phase space integrals are different
because of the $\epsilon$ dependence
 of $\alpha_2$ and $\beta_2$. Collecting 
the four contributions gives the average current
\begin{eqnarray}
\langle I \rangle\, &=& \, 2 \, \int 
\Bigl\lbrack \alpha_2^2(\epsilon) \, (2 e) \, \Gamma_{\alpha}(\epsilon) m_{\alpha} \nonumber \\
 &+& \sum_{j=1}^3 \beta_2^2(\epsilon) e^*_j \, \Gamma_{\beta, j }(\epsilon) m_j 
\Bigr\rbrack
d \epsilon, 
\end{eqnarray}
where the factor $2$ comes from the spin; 
$\Gamma_{\alpha}(\epsilon)$ and the $\Gamma_{\beta,j}$
 are $2 \pi/\hbar$ times the right mover self-energies
 due to the above mentionned processes and  $m$ is the multiplicity of the diagram.
 For instance, for the second process, $m_2=2$ and $e^*_2=4e$, so that 
the contribution is
$2 \,B^2 (\pi^2 \nu^2 T_K^2)^{-1}\,  m_2 e^* \,
 \nu^2 J$, with
 $J= \int_{- eV}^{eV} ({\epsilon /\Delta})^2 
(2 \pi / \hbar)\,  (1/4)\,
 \int_{- eV}^{\epsilon} dx \int_0^{\epsilon - x} dy \,\, d \epsilon$.
 For the first and third processes, $e^*_1=e^*_3=2e$ and $m_1 = m_2 = 4$, but
 the self-energy triple integrals obtained by integrating
 $\Gamma_{\beta,j}(\epsilon)$ on energy $\epsilon$ 
give a contribution $1/16$ smaller than
 for $({\cal R},{\cal R}) \rightarrow ({\cal L},{\cal L})$.

These results are summarized in Fig. 3,
showing the conductance $g=\langle I \rangle/V$ versus $eV/\Delta$, in units of
 $4 e^2/h$, for $\Delta/T_K = 0.2$. Many-body terms make $g$ decrease. 
In the case of $\Delta/T_K =0$, the BTK result \cite{BlonderTinkhamKlapwijk} is retrieved. 
For comparison, the result given by a resonant level formulation,
 without the many-body terms is shown.



{\it Noise:} We now turn to the zero frequency shot-noise calculation. As, to zero-th order
 in $\alpha_2$ and $\beta_2$, there is no partition noise (unlike SU(4) case$^{5,6}$), 
applying the Schottky formula for each process is sufficient$^4$. As in the normal case,
 direct inspection of the four processes 
 gives 
\begin{eqnarray}
S \, &=& \, 4 \, \int 
\Bigl\lbrack \alpha_2^2(\epsilon) \, (2 e)^2 \, \Gamma_{\alpha}(\epsilon) m_{\alpha}
\nonumber \\
 &+& \sum_{j=1}^3 \beta_2^2(\epsilon) (e^*_j)^2 \, \Gamma_{\beta, j }(\epsilon) m_j 
\Bigr\rbrack
d \epsilon. 
\end{eqnarray}
The resulting Fano factor with $A=B=2$ is thus $10/3$, to second order in $eV/T_K$ for any
 $eV \leq \Delta$. 


The finite frequency noise is now discussed. For a normal NS junction, at zero 
temperature, the absorption finite frequency noise $S(\omega)$ goes to zero for
 $\omega \, > \, 2 eV$. 
Here, within perturbation theory in Keldysh 
to second order in $1/T_K^2$, this is again the case. However,
 inspection of the diagrams suggests that inelastic terms bring more noise than in the case 
of an imperfect NS junction having the same Andreev conductance.


In conclusion, we have proposed a generalization of the floating of the Kondo resonance
 to the case where one electrode is superconducting, in the regime of
 small $\Delta/T_K$, close to the unitary limit. This enables a calculation 
of the many-body correction to the subgap Andreev current and zero frequency noise. 
Remarkably, the Fano factor sticks to the value $10/3$ as long as $eV \leq \Delta$. 
In the region $eV \ge \Delta$, not studied here, 
it is expected to decrease and to reach eventually
 the normal state limit $5/3$ for $eV \gg \Delta$.

\acknowledgments
M. Lavagna, J. L. Pichard and P. Simon are gratefully 
acknowledged for discussions and comments.

%
%
%
%
%
%
%
%
%
%
%
%
%
%
%
%

\begin{thebibliography}{99}
\bibitem{GlazmanPustilnik} 
L. I. Glazman and M. Pustilnik, in 
{\it Nanophysics: Coherence and Transport} edited by
 H. Bouchiat et al. (Elsevier 2005) pp. 427-478. 

\bibitem{Hewson} A. Hewson, {\it The Kondo Problem to Heavy Fermions} 
(Cambridge University Press, Cambridge 1993).

\bibitem{Nozieres} P. Nozi\`eres, J. Phys. Paris {\bf 39}, 1117 (1978).

\bibitem{SelavonOppen} E. Sela, Y. Oreg, F. von Oppen, and J. Koch, 
Phys. Rev. Lett. {\bf 97}, 086601 (2006). 

\bibitem{VitushinskyClerkleHur}
 P. Vitushinsky, A. A. Clerk, and K. Le Hur, Phys. Rev. Lett. 
{\bf 100}, 036603 (2008).

\bibitem{MoraLeyronasRegnault} C. Mora, X. Leyronas, and N. Regnault,
 Phys. Rev. Lett. {\bf 100}, 036604 (2008); 
{\it ibid.} {\bf 102}, 139902 (2009). 

\bibitem{Heiblumgroup} O. Zarchin, M. Zaffalon, M. Heiblum, D. Mahalu, 
and V. Umansky, Phys. Rev. B {\bf 77}, 241303(R) (2008).

\bibitem{Anderson}
P.W. Anderson, Phys. Rev. B {\bf 78}, 174505 (2008).

\bibitem{YoshiokaOhashi} T. Yoshioka and Y. Ohashi, 
J. Phys. Soc. Japan {\bf 69}, 1812 (2000).

\bibitem{AvishaiGZ} 
Y. Avishai, A. Golub, and A.D. Zaikin, Phys. Rev. B {\bf 63}, 134515 (2001);
 Y. Meir and A. Golub, Phys. Rev. Lett. {\bf 88}, 116802 (2002);
 Y. Avishai, A. Golub, and A. D. Zaikin, Phys. Rev. B {\bf 67}, 041301(R) (2003).

\bibitem{KangChoiBelzig} M. S. Choi, M. Lee, K. Kang, and W. Belzig, 
Phys. Rev. B {\bf 70}, 020502(R) (2004); Phys. Rev. Lett. {\bf 94}, 
229701 (2005). 

\bibitem{SianoEgger}
F. Siano and R. Egger, Phys. Rev. Lett. {\bf 93}, 047002 (2004), {\it ibid.}
 {\bf 94}, 229702 (2005).

\bibitem{LevyYeyati}
A. Levy Yeyati, A Mart{\`\i}n-Rodero,
 and E. Vecino, Phys. Rev. Lett. {\bf 91}, 266802 (2003).

\bibitem{Meden}
C. Karrasch, A. Oguri, and V. Meden, Phys. Rev. B {\bf 77}, 024517 (2008).

\bibitem{MLeeJonckMart}
M. Lee, T. Jonckheere, and T. Martin, Phys. Rev. Lett. {\bf 101}, 146804 (2008).

\bibitem{Wernsdorfergroup}
J.P. Cleuziou, W. Wernsdorfer, V. Bouchiat, T. Ondar{\c c}uhu, and M. Monthioux, 
Nature Nanotechnology {\bf 1}, 53 (2006).

\bibitem{DeblockBouchiat}
A. Eichler, R. Deblock, M. Weiss, C. Karrasch, V. Meden, C. Sch\"onenberger, and  H. Bouchiat, 
Phys. Rev. B {\bf 79}, 161407(R) (2009).

\bibitem{Beenakker}
C.W.J. Beenakker, Phys. Rev. B {\bf 46}, 12841 (1992).

\bibitem{BlonderTinkhamKlapwijk}
G. E. Blonder, M. Tinkham, and T. M. Klapwijk, 
Phys. Rev. B {\bf 25}, 4515 (1982).
\end{thebibliography}
\end{document}